# Unveiling Competition Dynamics in Mobile App Markets through User Reviews

Quim Motger[1[0000–0002–4896–7515]], Xavier Franch[1[0000–0001–9733–8830]], Vincenzo Gervasi[2[0000–0002–8567–9328]], and Jordi Marco[1[0000–0002–0078–7929]]

[1] Universitat Politècnica de Catalunya
{joaquim.motger,xavier.franch,jordi.marco}@upc.edu
[2] Università di Pisa
vincenzo.gervasi@unipi.it

**Abstract.** [**Context and motivation**] User reviews published in mobile app repositories are essential for understanding user satisfaction and engagement within a specific market segment. [**Question/problem**] Manual analysis of reviews is impractical due to the large data volume, and automated analysis faces challenges like data synthesis and reporting. This complicates the task for app providers in identifying patterns and significant events, especially in assessing the influence of competitor apps. Furthermore, review-based research is mostly limited to a single app or a single app provider, excluding potential competition analysis. Consequently, there is an open research challenge in leveraging user reviews to support cross-app analysis within a specific market segment. [**Principal ideas/results**] Following a case-study research method in the microblogging app market, we introduce an automatic, novel approach to support mobile app market analysis. Our approach leverages quantitative metrics and event detection techniques based on newly published user reviews. Significant events are proactively identified and summarized by comparing metric deviations with historical baseline indicators within the lifecycle of a mobile app. [**Contribution**] Results from our case study show empirical evidence of the detection of relevant events within the selected market segment, including software- or release-based events, contextual events and the emergence of new competitors.

## 1 Introduction

User reviews play a crucial role in providing valuable feedback and shaping the reputation of mobile apps [5]. They offer a wealth of information, including user satisfaction, engagement, and sentiments towards different aspects of the app [4]. Analysis of user reviews can uncover valuable insights into app performance and user preferences, and even identify emerging trends and potential issues [2,6,13]. However, manual analysis of these reviews is time-consuming, subjective, and impractical due to the informal nature and sheer volume of data [9]. As a result, app providers struggle to uncover hidden patterns, identify significant events, and understand the factors driving user satisfaction or dissatisfaction, especially if these emerge from a comparison with other similar and competing apps.

In fact, attention in research has been mostly devoted to internal analysis within the scope of a given mobile app or the catalogue of apps of a given app provider. We hypothesize that by leveraging a combination of quantitative metrics and event detection techniques, app providers can acquire a deeper and more timely understanding of a given app market. This entails any change in users' expectations and in the market in which they are competing, which may constitute a threat or present new opportunities to their business. This knowledge can then potentially support informed decisions that will materialize into new or evolved requirements. To the best of our knowledge, no other proposals exist for leveraging review-based metrics for cross-app and competition analysis within a given app market.

In this paper, we propose an automated, novel approach to unveil competition dynamics and explore mobile app market insights through the continuous analysis of user reviews and the detection of app-related events[3]. We collect user reviews from app repositories and utilize various well-established review-based metrics in the field to quantitatively assess user activity, satisfaction and engagement. To identify significant events, we focus on detecting deviations from baseline metrics. Reported events are used to select a subset of reviews from potentially correlated events. These reviews are summarized and presented to app providers to support explainability of such events. This, in turn, helps inform them about potential threats and opportunities that may inform the requirements elicitation for the future evolution of their app. We conducted a preliminary validation using the microblogging apps market as a case study to explore the potential of the approach (Sect. 4) and assess its novelty (Sect. 6).

## 2 Research method

### 2.1 Goal and research question

We perform a field study using a case study methodology [14], in order to facilitate the analysis of user review activity within its natural context (i.e., mobile app repositories) by means of minimal intrusion, limited to data collection. Our goal is **to proactively inform app stakeholders about changes in how users perceive a given app within a specific market segment, ultimately to infer the threats and opportunities stemming from these changes**. To this end, our research employs an exploratory design resulting in insights about the correlation between review-based metrics from a mobile app with respect to potential competitor apps. These insights focus on the automatic, proactive identification of significant events within a specific market segment and aim to monitor user behaviour and detect feedback trends. These events are then used to establish and detect potential user behavioural changes triggered by new market trends. To this end, we define the following research question:

> **RQ.** How can user review-based metrics be leveraged to uncover threats and opportunities from the user review activity of different mobile apps within the same market segment?

[3] Full datasets and complete evaluation results are available in the replication package: `https://doi.org/10.5281/zenodo.10125307`. Source code is also available at: `https://github.com/quim-motger/app-market-analysis`.

### 2.2 Stakeholder analysis

We identify three mobile app stakeholders that may benefit from our research through different types of analysis:

- **Providers**. Entities or individuals involved in the lifecycle of mobile apps, such as product owners, developers, marketers, and quality assurance teams. They might benefit by gaining insights into their app's performance and the competitive landscape, including identification of competitors, user trends, market threats and opportunities, aiding in decision-making regarding requirements, release planning, features, and niche identification.
- **Consultants**. Professionals or experts who provide specialized guidance to app providers, helping them optimize their strategies in a given app market. They can conduct in-depth analyses, comparing multiple apps' performance, features, and user feedback within a given market segment. This analysis provides valuable insights to identify market gaps, understand the positioning of different app providers, and propose informed decisions to stay competitive.
- **Users**. Individuals who interact with mobile apps as consumers. Our approach allows users to dynamically compare apps, considering evolving feedback trends and market dynamics. This empowers users to make more informed choices when selecting apps, potentially leading them to discover new and suitable candidates for their needs.

### 2.3 Case study: The "Twexit"

As preliminary evaluation for this paper, we selected an instrumental case study offering a contemporary, widely-discussed, and highly polarized event — the acquisition of X (formerly known as Twitter) by Elon Musk [10]. On April 14th, 2022, business magnate Elon Musk made an offer to purchase Twitter, Inc., owner of Twitter, one of the most popular microblogging apps. After several months of uncertainty, the acquisition was completed on October 27th, with opinions deeply divided between supporters and detractors of the operation. While some opinions praised Musk's stance on freedom of speech and no-censorship policies for Twitter, detractors reported great dissatisfaction with the upcoming changes in the Twitter app. This caused disruptions in the microblogging apps market for several months, leading to the migration of a significant amount of both users and advertisers to alternative existing apps, or even to the emergence of new competitors [10]. A popular example is the migration of Twitter users to Mastodon[4], an open-source, decentralized microblogging app, to which Twitter Inc. even reacted by banning Mastodon's accounts on their platform.

We chose to focus on the "Twexit" event for the following reasons:

– **Novelty and relevance:** Being a timely and widely discussed phenomenon within the microblogging app market, which makes it an ideal case to test our approach's ability to capture significant events and their consequences.
– **Real-world market dynamics:** Reflecting common market dynamics where sudden shifts in user perceptions and preferences occur due to high-profile events. Analysing this case provides valuable insights for adapting to evolving mobile app markets.
– **Assessment of generalization:** Serving as a well-known market phenomenon, it also provides a foundation for uncovering less obvious correlations and less popular events in the microblogging app market.

## 3 Approach

Our approach is composed of four stages (see Fig. 1): review collection, metric extraction, statistical analysis and event summarization. Each stage is described in some detail in the following subsections.

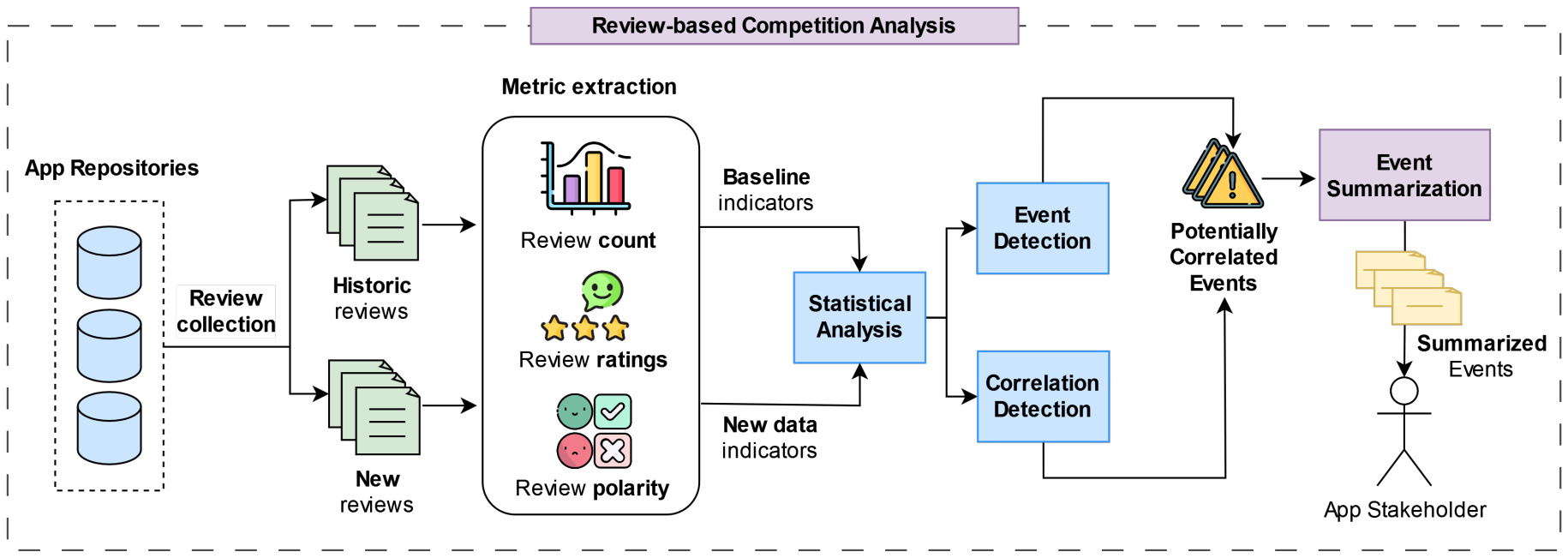

**Fig. 1.** System design

### 3.1 Review collection

Our approach combines web scraping and API consumption techniques developed in previous work [11]. This enables us to gather data from various decentralized and heterogeneous sources, including app stores, search engines, and recommendation platforms. We obtain a diverse range of information that includes both historic reviews and a periodic polling of new reviews for multiple apps in parallel in a given market segment. *Historic reviews* serve as a baseline capturing the app's typical activity patterns, enabling us to assess and compare the app's performance over time (see Sect. 3.3). Deviations from these established patterns can be used as robust signals for identifying significant market events. *New reviews* are collected periodically with configurable frequency (e.g., daily, weekly), serving as fresh data points to assess the current state of an app by comparing it with baseline statistical indicators.

[4] `https://joinmastodon.org/about`

### 3.2 Metric extraction

To support the event monitoring process, we focus on three established metrics that offer quantitative measures to assess changes in user engagement. This selection is motivated by the analysis of related work (see Section 6).

- **Review count** (c)**.** The number of published reviews is an indicator for monitoring disruptive events in the app market. Sudden surges or declines in review numbers can relate to multiple disruptions, such as emerging trends, market saturation, or shifts in user preferences.
- **Review rating** (r)**.** Ratings provide insights into user satisfaction and engagement. A significant increase may signal effective feature adoption, while a notable drop may indicate disruptive events like critical bugs or poorly received updates. Ratings are normalized to an integer value within $[0, 4]$.
- **Review polarity** (p)**.** Polarity measures opinion divergence among users. Extreme reviews indicate intense debate or controversy within an app. Polarity extends review ratings by capturing sentiment-based analysis at the sentence level, addressing limitations and capturing previously overlooked knowledge. Polarity is normalized to an integer value within $[0, 4]$, where 0 and 4 represent extremely negative and positive reviews, respectively.

Metrics c and r are directly collected and measured from the original app repositories through the review polling process. To compute p, we build on the work of a third-party sentiment analysis tool[5] based on a RoBERTa-based model. Notice metrics are computed uniformly using all reviews obtained from each data source.

### 3.3 Statistical analysis

Table 1 presents the formalization of the analytic metrics for measuring potentially correlated events. Below, we describe how these metrics are employed.

[5] Available at: `https://github.com/AgustiGM/sa_filter_tool`

**Table 1.** Event monitoring analytics

| **Analytic metric** | **Description** |
|---|---|
| Time Window $[t_0, t_0 + w)$ | Time window within the interval $[t_0, t_0 + w)$, where w is the interval length in number of days. |
| Average Value $\mu(m_{app}, [t_0, t_0 + w_e))$ | Average value of metric $m_{app}$ from all reviews published within the time window $[t_0, t_0 + w_e)$. |
| Average Difference $\delta(m_{app}, [t_0, t_0 + w_e))$ | Variation between $\mu(m_{app})$ and the average for the preceding time window $\mu(m, [t_0 - w_e, t_0))$. |
| Standard Deviation $\sigma(m_{app}, [t_0, t_0 + w_e), t_\Omega)$ | Baseline standard deviation for a given m measured from all $\delta(m_{app})$, where $t_\Omega \leq t_i < t_0$, being $t_\Omega$ a configurable date considered as the start date for baseline statistical computation. |
| Correlation $\rho(m_{app_i}, m_{app_j}, [t_0, t_0 + w_c), t_\Phi)$ | Pearson correlation between $m_{app_i}$ and $m_{app_j}$ measured from all $\mu(m_{app})$, where $t_\Phi \leq t_i < t_0$, being $t_\Phi$ a configurable date considered as the start date for baseline statistical computation. |

**Event detection.** Individual app timelines are used to monitor, detect and report events based on statistically significant deviations for a given metric. We use the complete set of reviews of a mobile app in a given market segment $app \in Apps$ to measure the average difference $\delta$ and the standard deviation $\sigma$ for each time window $[t_0, t_0 + w)$, as depicted in Table 1. Then, events for a given metric $m_{app} \in \{c, r, p\}$ are computed using the following formula:

$$E(m_{app}, t_0, w_e) = \begin{cases} 1 & \text{if } a \geq k \cdot s \\ 0 & \text{if} -k \cdot s < a < k \cdot s \\ -1 & \text{if } a \leq -k \cdot s \end{cases}$$

where $a = \delta((m_{app}, [t_0, t_0 + w_e))$, $s = \sigma((m_{app}, [t_0, t_0 + w_e))$ and k is a positive scalar determining the range of fluctuations that are considered normal: lower values of k lead to higher sensitiveness and more events being reported as significant; conversely, with higher values of k only fewer events of great momentum are reported. A new batch of user reviews collected during a specific time window $[t_0, t_0 + w_e)$ for a catalogue of mobile apps is processed to collect the aforementioned metrics and compute the monitoring analytics listed in Table 1. After this data collection and metric computation stage, we compute for each app in the identified market segment the value of $E(m_{app}, t_0, w_e)$.

**Correlation detection.** The event detection analysis is complemented with the statistical assessment of periods of time for which two mobile apps experienced a high (positive or negative) correlation (see Fig. 2). To this end, for each time window $[t_0, t_0 + w_c)$ and each pair of metrics $m_{app_i}, m_{app_j}$ where $app_i, app_j \in Apps$ and $i \neq j$, we measure the Pearson correlation $\rho(m_{app_i}, m_{app_j}, [t_0, t_0 + w_c))$ utilizing the set of reviews of $app_i, app_j$. Each correlation value is computed using all metric data points in $m_{app_i}, m_{app_j}$ for $[t_i, t_i + w_c) \mid t_{\Phi} \leq t_i < t_0 + w_c$, where $t_{\Phi}$ is the beginning of the observation period for correlation analysis. Correlated periods between $m_{app_i}$ and $m_{app_j}$ are then computed using the following formula:

$$C(m_{app_i}, m_{app_j}, t_0, w_c) = \begin{cases} 1 & \text{if } r \geq h \\ 0 & \text{if} -h < r < h \\ -1 & \text{if } r \leq -h \end{cases}$$

where $r = \rho(m_{app_i}, m_{app_j}, [t_0, t_0 + w_c))$ and $h \in [-1.0, 1.0]$ is defined as the sensitivity threshold for detecting correlated trends. Similarly to k, lower h values lead to a high sensitiveness for reporting correlated periods, while higher h values lead to a more strict detection. We employ a distinct time window $w_c$ to enhance our ability to detect correlations with a finer granularity. This enables us to identify even subtle or short-lived trends within the data more effectively.

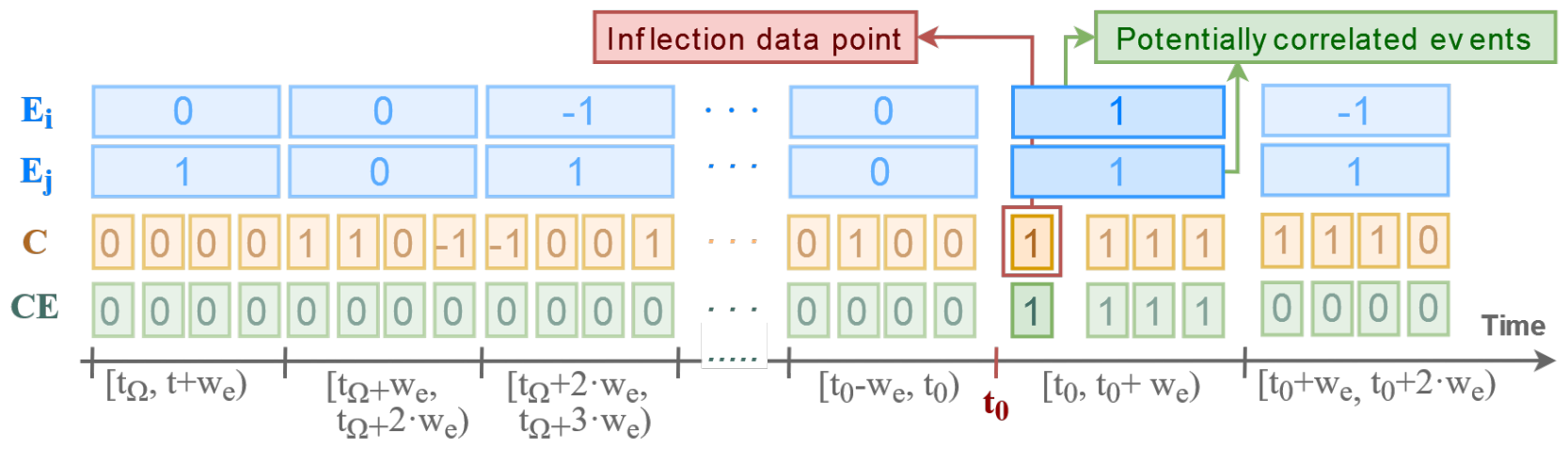

**Fig. 2.** Illustrative summary of the potentially correlated event detection method.

**Potentially correlated events detection.** Events and metric-based correlated periods are then used to find intersections of time windows $[t_0, t_0 + w_e)$ and $[t_0, t_0 + w_c)$ for which events were found at the same time in which there was a high correlation period (see Fig. 2). Hence, we define the following formula:

$$CE(m_{app_i}, m_{app_j}, t_0, w_e, w_c) = \begin{cases} 1 & \text{if } E_i = E_j \text{ and } C = 1 \\ -1 & \text{if } E_i \neq E_j \text{ and } C = -1 \\ 0 & \text{otherwise} \end{cases}$$

where we use the abbreviations $E_i = E(m_{app_i}, t_0, w_e)$, $E_j = E(m_{app_j}, t_0, w_e)$, and $C = C(m_{app_i}, m_{app_j}, t_0, w_c)$. Given the longest possible interval $[t_{start}, t_{end})$ from each period for which correlation detection is constantly $C = \pm 1$, we limit our analysis to intersections with respect to the first date interval $[t_{start}, t_{start} + w_e)$. This is motivated by the following reasons: (1) to focus exclusively on inflection data points of C, reflecting a change in the dynamic of the correlation $\rho$; and (2) to make our approach suitable for proactive detection in real-time, excluding future data from the analysis at each time window $[t_0, t_0 + w_e)$. Furthermore, since potentially correlated events may not always occur simultaneously (especially for lower $w_e$ values), we extend the $E_i = E_j$ assessment when $CE(m_{app_i}, m_{app_j}, t_0, w_e, w_c) = 0$ for the original time window $[t_0, t_0 + w_e)$. This extension includes the immediately preceding time window $[t_0 - w_e, t_0)$, evaluating $E_i = E_j$ for all date interval permutations within $\{[t_0 - w_e, t_0), [t_0, t_0 + w_e)\}$.

### 3.4 Event summarization

If $CE(m_{app_i}, m_{app_j}, t_0, w_e, w_c) \neq 0$, then our approach has detected potentially correlated events within $[t_0, t_0 + w_e)$. To support efficient interpretation, we complement our approach with a review summarization stage to make events suitable for human assessment. To this end, we employ ChatGPT (gpt3.5-turbo model) using prompt engineering[6] within a zero-shot learning approach to summarize a random sample of n reviews published within the time window $[t_0, t_0 + w)$. Each reported event $E_i, E_j$ corresponds to a particular summarization of the selected sub set of reviews associated with $E_i, E_j$. To illustrate this, in Summarization #1 we report the summarized output for an event triggered by the reviews of the Hive Social app (ellipsis added for presentation purposes).

**Summarization #1 - $E(c_{HiveSocial}, \text{“Nov 17, 2022”}, 7) = 1$** all

> The most significant event raised is the inability of users to create accounts on the Hive Social app. Many users reported encountering issues during the sign-up process, whether it was with email registration, phone number registration, [...]

In addition to generic summarization, our approach leverages the polarity metric p to split the reviews at the sentence level between negative sentences $(p \leq 1)$ and positive sentences $(p \geq 3)$. Each category is subsequently subjected to the same summarization process described above, enabling stakeholders to receive distinct summaries of diverse user perspectives on a specific event.

The following examples are the summarized output for positive and negative review sentences for the same event reported in Summarization #1. This example demonstrates the value of incorporating polarity-based distinction to extend the insights derived from the event summarization task. The summary of positive reviews (Summarization #2) allows deeper understanding of the underlying reasons for disruptive user activity in Hive Social, which might have been overlooked in the original summarization. Conversely, the summary of negative reviews (Summarization #3) offers a more nuanced analysis of the factors contributing to performance and functionality issues beyond user registration.

[6] Prompt template available in replication package.

Summarization #2 - $E(c_{\text{HiveSocial}}, \text{"Nov 17, 2022"}, 7)) = 1$ / $p \geq 3$ positive

The most significant event raised is the introduction and positive reception of a new app called "Hive." The reviewers express their initial impressions and enthusiasm for the app, highlighting its potential and features reminiscent of popular platforms like MySpace, [...] and Twitter. [...]

Summarization #3 - $E(c_{\text{HiveSocial}}, \text{"Nov 17, 2022"}, 7) = 1$ / $p \leq 1$ negative

The most significant event is the poor performance and functionality of the Hive social app, especially on Android devices. Many users reported issues such as slow loading times, frequent crashes, inability to upload or access photos, difficulties in creating accounts, and problems with basic features [...]

## 4 Evaluation

### 4.1 Design

We conducted a preliminary evaluation based on the validation of threats and opportunities as defined in Sect. 2.1 and the case study described in Sect. 2.3.

- **Data collection.** We employ the "Twexit" event as an exemplar to evaluate the effectiveness of our approach in identifying and reporting significant events within the microblogging app environment. To this end, we selected 12 microblogging apps for Android (including Twitter and Mastodon) based on users' recommendations from AlternativeTo[7]. For each app, we collected all reviews available in multiple repositories (Sect. 3.1) published within a time window of 52 weeks from June 9th, 2022 to June 7th, 2023 (included). Two of the apps were then excluded due to insufficient data (i.e., less than 20 reviews per month on average). The 10 remaining apps and corresponding number of reviews (187.639 in total) are listed in Fig. 3.
- **Parameter setup.** For event detection (E), we set $t_{\Omega}$ = "Jun 09, 2022" as the start date for baseline statistical computation; $w_e = 7$ as a weekly interval length for time window generation; and $k = \pm 2$ as the sensitivity factor. For correlation detection (C), we set $t_{\Phi} = t_0 - 14$ as start date for using recent data points (i.e., last 14 days); $w_c = 1$ for a fine-grained correlation analysis on a daily level; and $h = \pm 0.5$ as the sensitivity threshold value. Finally, for each potentially correlated event ($CE = \pm 1$), we performed event summarization on $n = 50$ as the average value of reviews published within a time window length of $w = 7$ is $\sim 41$, excluding Twitter (if the number of reviews in a given time window is $< 50$, we use all available documents).
- **Result analysis.** Evaluation is focused on: (1) verification of the ground truth case (i.e., "Twexit"); and (2) validation of additional potentially correlation events. To this end, we report statistics on potentially correlated events linked to the "Twexit", as well as complementary events that reflect other non-related competition phenomenon. In addition, event summaries are then categorized using a set of thematic categories of reasons to leave and join an app reported by actual users [1]. These categories include reasons related to usefulness, usability, content, reliability, security, existence of better alternatives, influence of others, popularity and design. We use these results to infer generic types of triggers of user trialling behaviours, supported by the examples and metrics selected in this study.

[7] From https://alternativeto.net/software/twitter/?platform=android

**Table 2.** Number of potentially correlated events detected for each metric $m = c, r, p$.

| | Events (E) | | | Correlations (C) | | | Correlated events (CE) | | |
|---|---|---|---|---|---|---|---|---|---|
| | +1 | −1 | ±1 | +1 | −1 | ±1 | +1 | −1 | ±1 |
| **count** (c) | 17 | 7 | 24 | 547 | 867 | 1,414 | 13 | 13 | 26 |
| **rating** (r) | 16 | 14 | 30 | 438 | 772 | 1,210 | 2 | 4 | 6 |
| **polarity** (p) | 27 | 23 | 50 | 309 | 717 | 1,026 | 1 | 2 | 3 |
| **Total** | 60 | 44 | 104 | 647 | 1,178 | 1,825 | 16 | 19 | 35 |

### 4.2 Results

Table 2 reports the summarized results for events (E), correlations (C) and potentially correlated events (CE) detection. Complementarily, Fig. 3 reports the distribution of events for the data set collected for the case study introduced in Sect. 2.3. From the original set of 104 events with $E = \pm 1$ for some metric $m_{app} \in \{c, r, p\}$, a correlation-based analysis reduces the scope of analysis to 35 potentially correlated event pairs $E_i, E_j$ covering 29 unique events with $CE = \pm 1$. Within the observation period (i.e., one year), from 35 pairs of potentially correlated events $E_i$ and $E_j$, 25 pairs (71.4%) are scoped within a 4-week span after the Twexit on October 27th, the trigger event of our ground truth. These 25 event pairs encompass 14 out of the 29 (48.3%) unique events E found in the set of potentially correlated events, and it affects 6 out of the 10 apps in our dataset, including Twitter and Mastodon (BlueSky was released a few months later). Among these, potentially correlated events include[8]: the migration of users from Twitter to Mastodon as a preferred alternative to some users (Summarization #4); the popularity of TruthSocial as a consolidated competitor motivated by the influence of public figures like Donald Trump (Summarization #5); and the emergence of Hive Social, a new competitor (Summarization #1).

**Summarization #4 - $E(c_{Mastodon}, \text{“Oct 27, 2022”}, 7) = 1$** — all

The most significant event [...] is the mixed reception and criticism of a Twitter alternative, likely referring to the Mastodon social media platform. [...]

[8] Complete summaries for all potentially correlated events are available in the replication package.

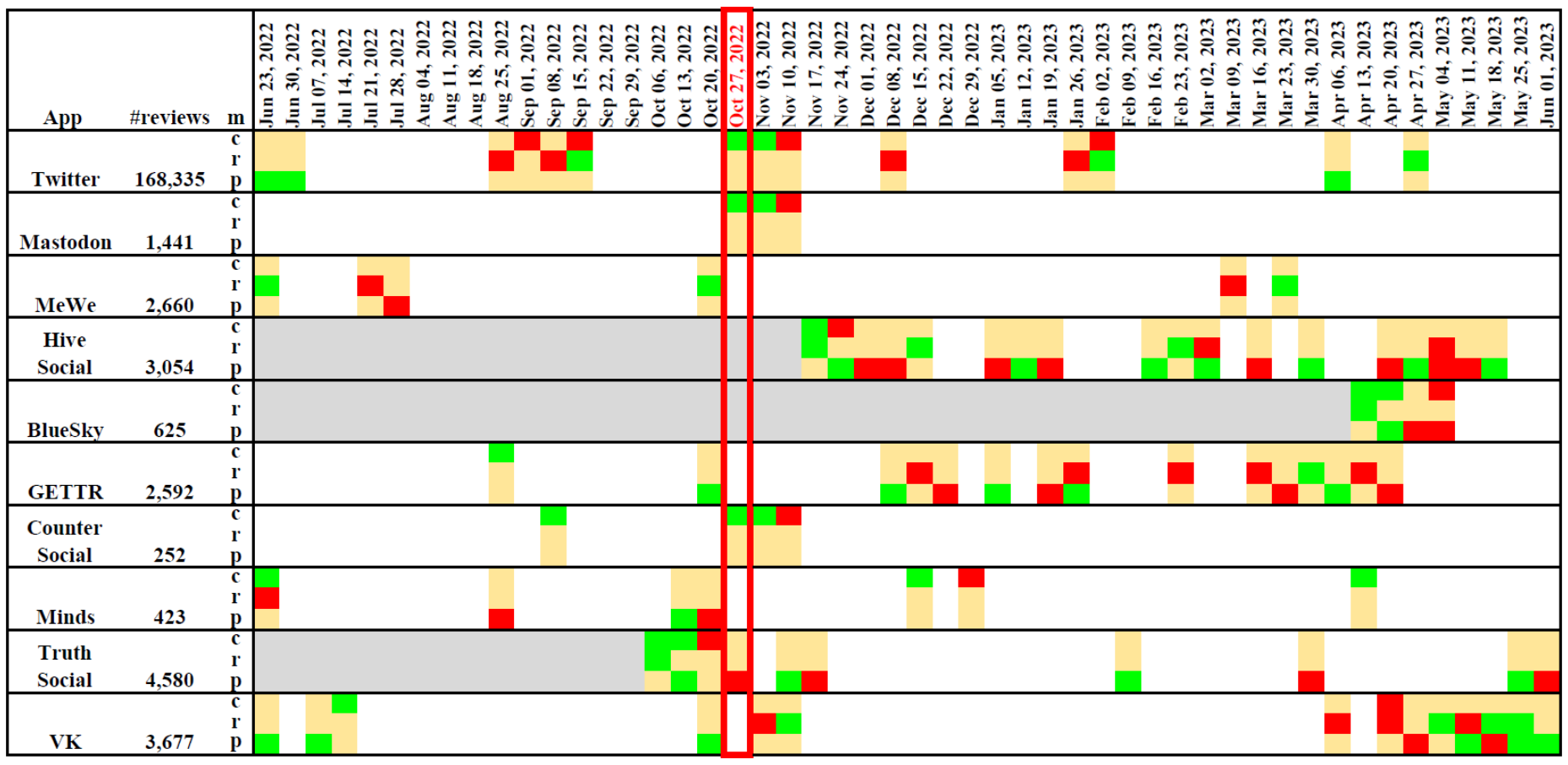

**Fig. 3.** Distribution of detected events for each app and each metric $m \in \{c, r, p\}$. Green and red cells represent positive and negative deviations for which $E(m_{app}, t_0, w_e) = \pm 1$. Yellow cells represent $E(m_{app}, t_0, w_e) = 0$ assessments for which another $m\prime$ reported $E(m\prime, t_0, w_e) = \pm 1$ for that same app and $[t_0, t_0 + w_e)$. Grey cells represent lack of data. The red box highlights the time window containing the date of the trigger event for the case study in Sect. 2.3 (i.e., acquisition of Twitter by Elon Musk).

**Summarization #5 - $E(c_{TruthSocial}, \text{“Nov 10, 2022”}, 7) = 1$ / $p \geq 3$** positive

The most significant event [...] is the launch of a new social media app called “Truth Social”, which is associated with former President Donald Trump. [...]

To extend the analysis to all potentially correlated events, Fig. 3 reports the results on the thematic-based classification of reasons to leave and join mobile apps by Al Shamaileh et al. [1]. We include the polarity-based distinction to identify how positive and negative reviews might infer different user perspectives in the summarization process. Frequency of each reason is consistent with results in the original study, being usability (24), usefulness (17) and content (16) the most frequent aspects mentioned in summaries. Filtering on positive reviews showcases a great impact on discussing the app as a potential better alternative to another one (19), being Twitter the reference app to which an alternative is being considered in all scenarios. Furthermore, influence of public figures like Musk or Trump also becomes significantly mentioned, both for negative (6) and positive (9) summaries. From this analysis, and using a sample representation of reported events, generic types of potentially correlated events can be inferred. Below, we highlight the most prominent ones, while also providing some examples for which the used metric was strictly necessary to detect that specific event. To explore the potential of our approach to detect and explain events beyond the ground truth event (i.e., Twexit), these examples refer to additional, complementary events, motivated by the classification in Fig. 4.

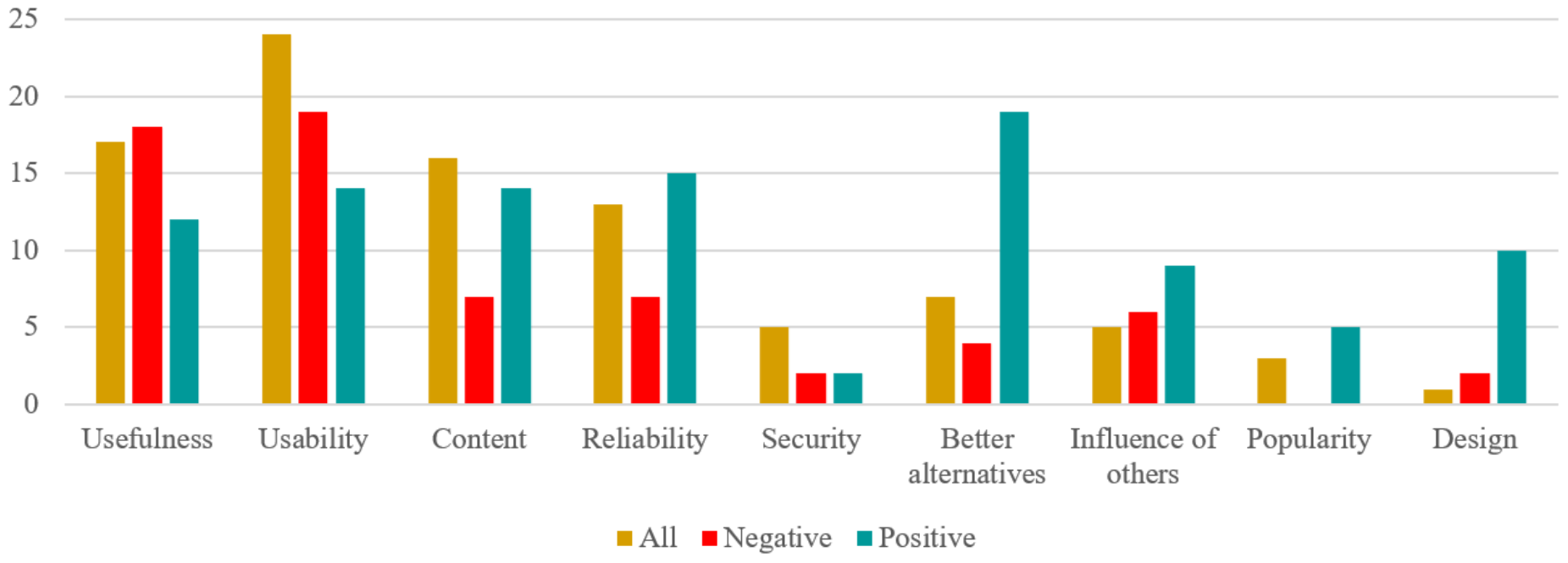

**Fig. 4.** Classification of unique events $E_i$ for which $CE = \pm 1$ using thematic categories of reasons to leave and join mobile apps as reported by Al Shamaileh et al. [1]

– **Software-based events.** Changes in the app's functionality, performance, updates, and user interface. They involve modifications in the codebase, backend infrastructure, or frontend interface that impact the user experience, system stability, or overall performance of the app.

Summarization #6 - $E(r_{VK}, \text{“Apr 06, 2023”}, 7) = -1$ / $p \leq 1$ negative

[...] Users have reported various issues, including login problems, account recovery issues, problems with OTP (One-Time Password) verification, excessive advertising, issues with video playback [...]

– **Context-based events.** External factors such as market trends, user preferences, regulations, or technological advancements that influence the app's usage, adoption, or user behavior. They highlight the dynamic interplay between the app and its surrounding environment.

Summarization #7 - $E(p_{Gettr}, \text{“Oct 20, 2022”}, 7) = 1$ all

The most significant event is the popularity of a social media app called Gettr [...] for various reasons, such as following and supporting a public figure named Andrew Tate, seeking an alternative to other social media platforms like Facebook and Twitter, [...]

– **New competitors.** Introduction of rival apps or platforms offering similar or alternative solutions in the same market space. It involves the entry of new players, their features, unique selling propositions, and potential impact on the user base, market share, and competitive landscape.

Summarization #8 - $E(c_{BlueSky}, \text{“Apr 20, 2023”}, 7) = 1$ / $p \geq 3$ positive

The most significant event [...] is the excitement and interest surrounding a new social media app called “BlueSky.” [...] They express positive sentiments about the app, stating that it's better than Twitter and that they prefer it.

# 5 Discussion

## 5.1 Findings

Our research question aims at identifying statistical analysis tasks (i.e., methods) that are useful for the automatic detection of significant threats and opportunities in a given market, which can be a source for new or evolved requirements. Evaluation results illustrate that the trigger event of the case study is actually reported by our approach. Furthermore, when examining the event detection results in Fig. 3 independently, it is not immediately evident that a disruptive pattern or event-based phenomenon related to the Twexit event is discernible. It is only when we augment the review-based analysis with a correlation analysis that we observe a significant concentration of events during the Twexit time frame. This observation underscores the enhanced utility of a cross-app statistical analysis. Moreover, beyond Twitter-to-Mastodon user migration, multiple examples of minor events which might not have such a major impact (e.g., new competitors, positive/negative reactions to new releases from competitors) are also proactively detected. Finally, it is essential to emphasize that our approach does not assess the actual cause-and-effect relationships between potentially correlated events. Instead, our focus is on detecting significant events that represent anomalies in a single app's timeline, which occur simultaneously in time. These events might ultimately provide insights into situations where users may be inclined to leave or join a particular app.

Beyond ground truth events, event summarization (as shown in Summarization #1–#8) and categorization (in Fig. 4) showcase the potential of our approach to support market analysis use cases (depicted in Sect. 2.2). An in-depth analysis of a well-known period of observation (i.e., one year) can serve app providers entering a new market or app consultants to conduct SWOT analysis on a given app's market and support informed decisions on strategic alignment, requirements elicitation and release planning. Moreover, metric analysis and methods are designed to support timely, up-to-date detection of events exclusively based on past historical data. Hence, app providers within a given market can use these insights to support continuous development, software maintenance and user feedback analysis. Finally, these results illustrate our approach under a specific set up for the defined configurable parameters (enumerated in Sect 4.1). Given that the significance of an event is a subjective and contextual notion, we do not seek to define optimal values for these parameters. Moreover, they could be set up differently at different times, according to stakeholder-specific criteria, and to the specific information need that has to be satisfied at any point in time.

## 5.2 Threats to validity

We examine our study's constraints by addressing the validity issues outlined by Wohlin et al. [16]. Concerning construct and internal validity, the accuracy of the statistical analysis relies heavily on the quantity of timely available data.

Continuous access to a sufficient amount of reviews is essential to avoid missing events. On the other hand, different review metrics and metric computation may lead to varying results. In this research, we have focused on the use of established metrics covered by related work (see Sect. 6). Examples in Sect. 4 illustrate multiple scenarios in which each metric $m_{app} \in \{c, r, p\}$ disjointly reports multiple events. These examples contribute to the hypothesis that the selection of metrics for our approach is appropriate to respond to the research question. Furthermore, we plan to extend the analysis to complementary review metrics as future work. Additionally, the use of specific parameters for the evaluation might introduce bias, as different parameter settings could yield varied results, potentially impacting the study's internal consistency. We do not claim that the selection of parameters answers to a specific validity of an optimal setting of our approach. As mentioned in Sect. 5.1, parameters must be properly selected according to stakeholders interests. Concerning event summarization, using a random sample of n reviews might introduce a bias in the summarization process. To this end, we plan to extend our approach by integrating an internal summarization process (see Section 7). Finally, the use of the Twexit event as ground-truth poses a generalization bias, which we attempted to mitigate by extending evaluation with the analysis of all potentially correlated events and the thematic-based classification of reasons to leave and join mobile apps.

Concerning external validity, we rely on gpt3.5-turbo model instance from OpenAI on an unsupervised setting for event summarization. This decision was made at this research stage to focus on the exploratory nature of our research question, based on validating potentially correlated events. We used gpt3.5-turbo based on its demonstrated accuracy on document summarization on unsupervised tasks [3]. We also rely on an external tool for computing the polarity (p) metric. Finally, concerning conclusion validity, main limitations emerge from generalization beyond the microblogging market or even beyond the set of apps and reviews in the dataset. To mitigate these limitations, we utilize a diverse dataset with various sources and reviews in the microblogging market from a large period of observation (i.e., 1 year), increasing the depth and breadth of our insights and minimizing potential generalization issues.

## 6 Related work

There is an increasing body of research on the analysis of user reviews in software engineering [5]. Most of this body of research focuses on user reviews for a single app, mainly with the purpose of extracting feature requests, bug reports or more generally, requirements for next releases. Similar to our approach, Gao et al. [8] analyse time slices corresponding to releases of a given app in order to detect emerging issues, namely bugs or unfavourable app features. Our concept of event is more general and includes other possible market events. Strønstad et al. [15] proposed a supervised machine learning-based pipeline for the automatic detection of anomalies in single-app time series, based on metrics related to (1) statistical data, (2) ratings and upvotes, (3) sentiment analysis, and (4) review content. However, being single-app oriented, they do not try to connect reviews from different competitive apps in order to get a holistic view of the marketplace.

In fact, approaches considering user reviews for multiple apps are scarce, and even less aim at assessing the position of an app in the market. Shah et al. [13] compare a reference app with its competitors in terms of sentiment, bug reports and feature requests. Dalpiaz and Parente [6] perform a SWOT analysis for a reference app after comparing with the sentiment of competitor apps as expressed in user reviews. Assi et al. [2] identify high-level features mentioned in user reviews and create a comparative table that summarizes users' opinions for each identified feature across competing apps. The three approaches provide snapshots of the marketplace in a given moment of time, without including the time dimension, as our approach does. While timely evaluation of user reviews has been explored before for single app analysis [7], our approach could be integrated with any of these methods, providing the missing time dimension in the context of whole-market segment analysis. In addition, these methods use classical feature extraction and topic modelling techniques at the feature/topic granularity level. Instead, we focus on the app as a whole and delegate onto ChatGPT3.5 the summarization of the cause of events. Finally, there are a few recent studies assessing the causes for app switching behaviours in the context of mobile app markets [1, 12]. However, these are limited to survey and interview-based methods, and do not refer to automated proposals for identifying or reporting these phenomenona.

## 7 Conclusion and future work

In this paper, we have presented a novel approach for automated event monitoring in a mobile app market segment. Evaluation results illustrate several timely and actionable examples regarding market dynamics, including software-level key changes, contextual factors, and the entry of new competitors. A systematic approach to continuously report these types of events could be applied in multiple use cases, including: app providers to enhance tasks such as requirements elicitation, prioritization and release planning; a consulting company to provide market consultancy in software and requirements engineering tasks; a large organization to decide which app to select for internal use to conduct a business activity. Moreover, the selected review metrics and statistical methods provide valuable insights to support the research hypothesis, while also opening the scope for future work. We plan on expanding the original metrics, evaluating their covariance structure, and exploring potential inner correlations. We will also enhance the selection mechanism and employ quality-based filtering techniques to improve the set of reviews used for summarization. Additionally, we aim to extend the reported knowledge and enhance the explainability of event monitoring results by analysing variations in outcomes across different experimentation setups. Last, we plan to combine our approach with some of the methods mentioned in the related work to combine our time dimension with the feature-oriented granularity that these methods provide.

## Acknowledgment

With the support from the Secretariat for Universities and Research of the Ministry of Business and Knowledge of the Government of Catalonia and the European Social Fund. This paper has been funded by the Spanish Ministerio de Ciencia e Innovación under project / funding scheme PID2020-117191RB-I00 / AEI/10.13039/501100011033.

## Replication package

The complementary replication package contains the following artefacts:

- **Dataset:** The dataset of reviews used for the evaluation of our approach.
- **Software Components:** Employed for metric extraction, event detection, correlation detection, and identification of potentially correlated events.
- **Metric Values:** The resulting metric values for c, r and p.
- **Statistical Analysis:** Detailed records of the identified events, correlations, and potentially correlated events for all pairwise combinations of c, r and p.
- **Summaries:** Complete summaries for all potentially correlated events.